\documentstyle[12pt]{article}
\newcommand\be{\begin{equation}}
\newcommand\ee{\end{equation}}
\newcommand\bea{\begin{eqnarray}}
\newcommand\eea{\end{eqnarray}}
\newcommand\ket[1]{|#1\rangle}

\newcommand{\fatalpha}{{\bf \alpha \kern -0.44em \alpha}}
\newcommand{\fatsigma}{{\bf \sigma \kern -0.54em \sigma}}
\newcommand{\tpchi}{{\bf \chi \kern -0.35em \chi}}
\newcommand{\llambda}{{\bf \lambda \kern -0.45em \lambda}}

\bibliography{plain}
\title{\bf Investigation graph isomorphism problem via entanglement entropy in strongly regular graphs}\vspace{20mm}
\author{ M. A. Jafarizadeh$^{a}$
 \thanks{E-mail:jafarizadeh@tabrizu.ac.ir},
 F. Eghbalifam$^{a}$
 \thanks{E-mail:F.Egbali@tabrizu.ac.ir},
 S. Nami$^{a}$
 \thanks{E-mail:S.Nami@tabrizu.ac.ir}
\\ $^a${\small Department of Theoretical Physics and Astrophysics,
University of Tabriz, Tabriz 51664, Iran.}} \pagebreak

\vspace{20mm}
\begin{document}
\maketitle \vspace{15mm}
\begin{abstract}
We investigate the quantum networks that their nodes are considered as quantum
harmonic oscillators. The entanglement of the ground state can be used to quantify the
amount of information one part of a network shares with the other part of the system.
The networks which we studied in this paper, are called strongly regular graphs (SRG).
These kinds of graphs have some special properties like they have three strata in the
stratification basis. The Schur complement method is used to calculate the Schmidt
number and entanglement entropy between two parts of graph. We could obtain analytically, all blocks of adjacency matrix in several important kinds of strongly regular
graphs. Also the entanglement entropy in the large coupling limit is considered in these
graphs and the relationship between Entanglement entropy and the ratio of size
of boundary to size of the system  is found. Then, area-law is studied to show that there are no entanglement entropy for the highest size of system.
\\Then, the graph isomorphism problem is considered in SRGs by using the elements of blocks of adjacency matrices. Two SRGs with the same parameters:$(n, \kappa,\lambda ,\nu)$
are isomorphic if they can be made identical by relabeling their vertices. So the adjacency matrices of two isomorphic SRGs become identical by replacing of rows and
columns. The nonisomirph SRGs  could be distinguished  by using the elements of blocks of
adjacency matrices in the stratification basis, numerically.
\end{abstract}
\section{Introduction}
Entanglement plays a crucial role in quantum information
processing, including quantum communication [1,2] and quantum
computation [3-5]. It is one of the remarkable features that
distinguishes quantum mechanics from classical mechanics.\\ For
decades, entanglement has been the focus of much work in the
foundations of quantum mechanics, being associated particularly
with quantum nonseparability and the violation of Bells
inequalities[6]. In recent years, however, it has begun to be
viewed also as a potentially useful resource. The predicted
capabilities of a quantum computer, for example, rely crucially on
entanglement[7]. \\The synergy between the field of complex
networks and that of information theory has recently appealed to
the quantum information community. The use of quantum dynamical
processes, such as quantum random walks [8] has given new
quantum information perspectives to classical problems of the
network realm.\\ In [9] the authors quantified the amount of
information that a single element of a quantum network shares with
the rest of the system. They considered a network of quantum
harmonic oscillators and analyzed its ground state to compute the
entropy of entanglement that vacuum fluctuations creates between
single nodes and the rest of the network by using the entropy of
entanglement, defined as the Von Neumann entropy. \\One of the
important problems about networks is the graph isomorphism
problem. Two graphs are isomorphic, if one can be transformed into
the other by a relabeling of vertices (i.e. two graphs with the
same number of vertices and edges are nonisomorph, if they can not
be transformed into each other by relabeling of vertices). Many
graph pairs may be distinguished by a classical algorithm which
runs in a time polynomial in the number of vertices of the graphs,
but there exist pairs which are computationally difficult to
distinguish. Currently, the best general classical algorithm has a
run time $O(c^{\sqrt{N}\log N})$, where c is a constant and N is the
number of vertices in the two graphs. Strongly regular graphs
(SRGs) are a particular class of graphs that have four dependent parameter $(n,\kappa,\lambda,\mu)$, that are dificult to
distinguish classically [10]. Graph isomorphism is believed to be similar to
factoring in that both are though to be NP-Intermediate problem
$[11]$. Additionally, both problems may be approached as hidden
subgroup problems, though this approach has had limited success
for GI$[25]$. Due to these similarities, and the known quantum
speedup available for factoring $[26]$, there is hope that there
similarly exists a quantum speedup for GI. \\Strongly regular
graphs (SRGs) are a particular class of graphs that are difficult
to distinguish classically. One class of algorithms that has been
explored for GI is that of quantum random walks. \\Shiau et al. showed that the
single-particle continuous-time QRW fails to distinguish pairs of
SRGs with the same family parameters [11]. Gamble et al. extended
these results, proving that QRWs of two noninteracting particles
will always fail to distinguish pairs of nonisomorphic SRGs with
the same family parameters [12].\\ Then Rudinger et al.
numerically demonstrated that three-particle noninteracting walks
have distinguishing power on pairs of SRGs [13,14].\\In this paper, we
want to investigate the graph isomorphism problem in strongly
regular graphs. To this aim, we use the entanglement to
distinguish two nonisomorph SRGs. So first we use the
stratification techniques [15-19], to write the adjacency matrices of SRGs
in the block form. The obtained matrix, becomes block diagonal in
the stratification basis. We called it the block-diagonal
adjacency matrix. The first block of obtained matrix, will be a $3
\times 3$ matrix and the other blocks are $2 \times 2$ or
singlets. The $3 \times 3$ block is related only to the parameters
of the SRG and obtains analytically in terms of parameters for all
SRGs. The entanglement entropy can be used for calculating the
entanglement between two parts of graph. The entanglement between
the first stratum (which has only one vertex) and other vertices
(second and third strata), will be obtained only from $3 \times3$
matrix. But for calculating the entanglement between other
subsets, we need the $3 \times 3$ and all of the $2 \times 2$
blocks of adjacency matrix. We discuss about the elements of these
$2 \times 2$ blocks and give some important relations between its
elements. Also for several important kinds of SRGs, we could
obtain the $2 \times 2$ blocks of adjacency matrices analytically.
So the entanglement entropy between all two subsets, will be
obtained in these kinds of SRGs analytically. For the other SRGs
which their adjacency matrices were identified, we could calculate
the block-diagonal adjacency matrix numerically and could
distinguish the nonisomorph SRGs from their $2 \times 2$ blocks.
\\In the section II, first we describe the Hamiltonian of our
model in subsection 2.1. Also we demonstrate the Schmidt
decomposition and entanglement entropy in 2.2. Finally in 2.3 we give
some properties of strongly regular graphs.\\ In
section III, we calculated the entanglement entropy between two
parts of the SRGs. It is performed by using the Schur complement
method and some local transformations. We obtained the Schmidt
decomposition and Schmidt numbers of the ground state wave
function.\\ In section IV, we
used the spectral techniques to obtain some important relations
for the elements of the block-diagonal adjacency matrix. \\In this section, we give some simple kinds of SRGs. These
kinds of SRGs don't contain nonisomorph graphs. Then we
give four kinds of SRGs in four examples that we obtain their
blockdiagonal adjacency matrices by using the relations of
previous section and the information of graphs analytically. For
each of examples we show that we can distinguish nonisomorph SRGs
from the block-diagonal adjacency matrix. \\In section V, we
give some other examples of nonisomorph SRGs which can be distinguished
by using their block-diagonal adjacency matrices numerically. The Schur complement method is in the Appendix A and
the stratification techniques are given in Appendix B.
\section{Preliminaries}
\subsection{The model and hamiltonian}
The nodes are considered as identical quantum oscillators, interacting as
dictated by the network topology encoded in the Laplacian $L$. The
Laplacian of a network is defined from the Adjacency matrix as
$L_{ij} = k_i\delta_{ij}- A_{ij}$ , where $k_i =\sum_j A_{ij}$ is
the connectivity of node $i$, i.e., the number of nodes connected
to $i$. The Hamiltonian of the quantum network thus reads:
\begin{equation}
H=\frac{1}{2}(P^T P+ X^T(I+2gL)X)
\end{equation}
here $I$ is the $N \times N$ identity matrix, $g$ is the coupling
strength between connected oscillators while $p^T=(p_1,p_2,...,
p_N)$ and $x^T=(x_1,x_2, ..., x_N)$ are the operators
corresponding to the momenta and positions of nodes respectively,
satisfying the usual commutation relations: $[x, p^T] = i\hbar I$
(we set $\hbar = 1$ in the following) and the matrix $V=I+2gL$ is
the potential matrix. Then the ground state of this Hamiltonian
is:
\begin{equation}
\psi(X)=\frac{(det(I+2gL))^{1/4}}{\pi^{N/4}}exp(-\frac{1}{2}(X^T(I+2gL)X))
\end{equation}
Where the $A_g=\frac{(det(I+2gL))^{1/4}}{\pi^{N/4}}$ is the
normalization factor for wave function. The elements of the
potential matrix in terms of entries of adjacency matrix is
$$V_{ij}=(1+2g\kappa_i)\delta_{ij}-2gA_{ij}$$
\subsection{Schmidt decomposition and entanglement entropy}
The Schmidt decomposition is a very good tool to study
entanglement of bipartite pure states. The Schmidt number provides
an important variable to classify entanglement. Any bipartite pure
state $|\psi\rangle_{AB} \in \textsl{H}=\textsl{H}_A
\otimes\textsl{H}_B$ can be decomposed, by choosing an appropriate
basis, as
\begin{equation}
|\psi\rangle_{AB}=\sum_{i=1}^m
\alpha_i|a_i\rangle\otimes|b_i\rangle
\end{equation}
where $1 \leq m \leq min\{dim(\textsl{H}_A); dim(\textsl{H}_B)\}$,
and $\alpha_i
> 0$ with $\sum_{i=1}^m \alpha_i^2 = 1$. Here $|a_i\rangle$ ($|b_i\rangle$) form a part of an
orthonormal basis in $\textsl{H}_A$ ($\textsl{H}_B$). The positive
numbers $\alpha_i$ are called the Schmidt coefficients of
$|\psi\rangle_{AB}$ and the number $m$ is called the Schmidt rank
of $|\psi\rangle_{AB}$. The entanglement of a partly entangled
pure state can be naturally parameterized by its entropy of
entanglement, defined as the von Neumann entropy of either
$\rho_A$ or $\rho_B$, or equivalently as the Shannon entropy of
the squares of the Schmidt coefficients [23].
\begin{equation}
E=-Tr\rho_A log_2\rho_A= Tr\rho_B log_2\rho_B=-\sum_i\alpha_i^2
log_2 \alpha_i^2
\end{equation}
\subsection{Strongly regular graphs(SRG)} A graph (simple,
A graph (simple, undirected
and loopless) of order $n$ is strongly regular with parameters
$n$, $\kappa$, $\lambda$, $\mu$ whenever it is not complete or
edgeless and

(i) each vertex is adjacent to $\kappa$ vertices,

(ii) for each pair of adjacent vertices there are $\lambda$
vertices adjacent to both,

(iii) for each pair of non-adjacent vertices there are $\mu$
vertices adjacent to both.

We assume throughout that a strongly regular graph $G$ is
connected and that $G$ is not a complete graph. Consequently,
$\kappa$ is an eigenvalue of the adjacency matrix of G with
multiplicity $1$ and
\begin{equation}
n-1>\kappa \geq\mu>0\quad,\quad \kappa-1>\lambda\geq0
\end{equation}
Counting the number of edges in $G$ connecting the vertices
adjacent to a vertex $x$ and the vertices not adjacent to $x$ in two
ways we obtain
\begin{equation}
\kappa(\kappa-\lambda-1) = (n-\kappa-1)\mu
\end{equation}
So the relation between these parameters is
\begin{equation}
\kappa^2 = (\kappa-\mu) +\mu n + (\lambda-\mu)\kappa
\end{equation}

The adjacency matrix of any SRG satisfies the particularly useful
algebraic identity
\begin{equation}
A^2 = (\kappa-\mu)I +\mu J + (\lambda-\mu)A
\end{equation}
where $I$ is the identity and $J$ is the matrix of all ones.

\section{Entropy of entanglement in the ground state of quantum harmonic oscillators}
In order to calculate the entanglement entropy between two
parts in the graph (for example strata $1$ and $(2,3)$), we
introduce the following process: First one divide the potential
matrix of the graph into three part as
\begin{equation}
V=I+2gL=\left(\begin{array}{ccc}
          V_{11}& V_{12} & 0\\
            V_{21}& V_{22} & V_{23}\\
            0 & V_{32} & V_{33}\\
          \end{array}\right)
\end{equation}

Then by using the generalized Schur complement method, the potential matrix can be write
$$\left(\begin{array}{ccc}
         V_{11}& V_{12} & 0 \\
            V_{21}& V_{22} & V_{23}\\
            0 & V_{32} & V_{33}\\
          \end{array}\right)=$$
\begin{equation}
          \left(\begin{array}{ccc}
          I_1& 0 & 0\\
            0& I_2& V_{23}V_{33}^{-1}\\
            0 & 0 & I_3\\
          \end{array}\right)\left(\begin{array}{ccc}
          V_{11}& V_{12} &0 \\
            V_{12}^{T}& V_{22}-V_{23}V_{33}^{-1}V_{32} & 0\\
            0 & 0 & V_{33}\\
          \end{array}\right)\left(\begin{array}{ccc}
          I_1& 0 & 0\\
            0& I_2& 0\\
            0 & V_{33}^{-1}V_{32} & I_3\\
          \end{array}\right)
\end{equation}
In the transformed matrix the blocks  are scalar. So for calculating the entanglement between two subsets, it is sufficient to use the $2\times 2$ matrix as
\begin{equation}
\left(\begin{array}{cc}
         a_{11}&a_{12} \\
          a^T_{12}& a_{22}\\
          \end{array}\right)=\left(\begin{array}{cc}
         V_{11}&V_{12} \\
            V_{12}^{T} & V_{22}-V_{23}V_{33}^{-1}V_{32}\\
          \end{array}\right)
\end{equation}
The wave function in this stage is
\begin{equation}
\psi(x,y)=A_g exp(-\frac{1}{2}(x\quad \quad
y)\left(\begin{array}{cc}
          a_{11}&a_{12}\\
            a_{12}& a_{22}\\
          \end{array}\right)\left(\begin{array}{c}
          x\\
            y\\
          \end{array}\right))
\end{equation}
by rescaling the variables $x$ and $y$:
$$\widetilde{x}=a_{11}^{1/2}x$$
$$\widetilde{y}=a_{22}^{1/2}y$$
the ground state wave function is transformed to
\begin{equation}
\psi(\widetilde{x},\widetilde{y})=A_g
exp(-\frac{1}{2}(\widetilde{x}\quad \quad
\widetilde{y})\left(\begin{array}{cc}
          1& d\\
            d& 1\\
          \end{array}\right)\left(\begin{array}{c}
          \widetilde{x}\\
            \widetilde{y}\\
          \end{array}\right))
\end{equation}
where $d=a_{11}^{-1/2}a_{12}a_{22}^{-1/2}$. So the ground state
wave function is
\begin{equation}
\psi(\widetilde{x},\widetilde{y})=A_g
e^{-\frac{\widetilde{x}^2}{2}-\frac{\widetilde{y}^2}{2}-d\widetilde{x}\widetilde{y}}
\end{equation}

From above equation, it's clear that the node $\widetilde{x}$ is
just entangled with $\widetilde{y}$, so one can use following
identity to calculate the schmidt number of this wave function,
\begin{equation}
\frac{1}{\pi^{1/2}}exp(-\frac{1+t^2}{2(1-t^2)}((\widetilde{x})^2+(\widetilde{y})^2))+\frac{2t}{1-t^2}\widetilde{x}
\widetilde{y})=(1-t^2)^{1/2}\sum_n t^n
\psi_n(\widetilde{x})\psi_n(\widetilde{y})
\end{equation}
In order to calculating the entropy, we apply a change of variable
as
$$1-t^2=\frac{2}{\gamma+1}$$
$$t^2=\frac{\gamma-1}{\gamma+1}$$
So the above identity becomes
\begin{equation}
\frac{1}{\pi^{1/2}}exp(-\frac{\gamma}{2}((\widetilde{x})^2+(\widetilde{y})^2))+(\gamma^2-1)^{1/2}\widetilde{x}
\widetilde{y})=(\frac{2}{\gamma+1})^{1/2}\sum_n
(\frac{\gamma-1}{\gamma+1})^{n/2}
\psi_n(\widetilde{x})\psi_n(\widetilde{y})
\end{equation}
and the reduced density matrix is
\begin{equation}
\rho=\frac{2}{\gamma+1}\sum_{n}(\frac{\gamma-1}{\gamma+1})^n
|n\rangle \langle n|
\end{equation}
and the entropy is
\begin{equation}
S(\rho)=\frac{\gamma +1}{2} log(\frac{\gamma +1}{2})-\frac{\gamma
-1}{2}log(\frac{\gamma -1}{2})
\end{equation}
By definition the scale $\mu^2$, we obtain
$$\gamma=1 \times \mu^2$$
$$(\gamma^2-1)^{1/2}=-d \times \mu^2$$
After some straightforward calculation
\begin{equation}
\gamma=(\frac{1}{1-d^2})^{1/2}
\end{equation}
Where $d$ is schmidt number.
\section{Calculating Bipartite entanglement in stratificatin basis of SRG}
The adjacency matrix for a strongly regular graph is
\begin{equation}
A=|0\rangle\langle1|\otimes
e_{\kappa}^{T}+|1\rangle\langle0|\otimes
e_{\kappa}+|1\rangle\langle1|\otimes
A_{11}+|1\rangle\langle2|\otimes A_{12}+|2\rangle\langle1|\otimes
A_{12}^{T}+|2\rangle\langle2|\otimes A_{22}
\end{equation}
And
\begin{equation}
A^2=\left(\begin{array}{ccc}
          k & e_{\kappa}^T A_{11} & e_{\kappa}^T A_{12}\\
            A_{11}e_{\kappa}& e_\kappa e_\kappa^T+A_{11}^2+A_{12}A_{12}^T & A_{11}A_{12}+A_{12}A_{22}\\
            A_{12}^Te_{\kappa} & A_{12}^TA_{11}+A_{22}A_{12}^T &A_{12}^TA_{12}+A_{22}^2 \\
          \end{array}\right)
\end{equation}
From the block (1,3) of equation (4-20) and  equation (2-8) we
conclude that
$$e_{\kappa}^T A_{12}=\mu e_{\kappa'}^T$$
\begin{equation}
A_{12}^T e_\kappa=\mu e_{\kappa'}
\end{equation}
So
\begin{equation}
\sum_{\alpha}(A_{12})_{\alpha ,j}=\mu
\end{equation}
Also can be written from the block (1,2)
\begin{equation}
\sum_{\alpha}(A_{11})_{\alpha j}=\lambda,\quad\quad
\sum_{\alpha}(A_{11})_{j \alpha}=\lambda
\end{equation}
From the block (2,3) of $A^2$ we have
\begin{equation}
A_{11}A_{12}+A_{12}A_{22}=\mu J_{\kappa\kappa'}+(\lambda
-\mu)A_{12}
\end{equation}
Then we multiply the above equation from the left side in
$e_\kappa^T$ and use the equations (4-21) and (4-22) to prove that
\begin{equation}
\sum_{\alpha}(A_{22})_{\alpha j}=\kappa-\mu,\quad\quad
\sum_{\alpha}(A_{22})_{j \alpha}=\kappa-\mu
\end{equation}
Other equations are
\begin{equation}
A_{12}^TA_{12}+A_{22}^2=(\kappa-\mu)I_{\kappa'}+\mu
J_{\kappa'\kappa'}+(\lambda-\mu)A_{22}
\end{equation}
\begin{equation}
A_{11}^2+A_{12}A_{12}^T=(\kappa-\mu)I_{\kappa}+(\mu-1)
J_{\kappa\kappa}+(\lambda-\mu)A_{11}
\end{equation}

Suppose that $A_{12}=O_1D_{12}O_2^T$ be the singular value
decomposition of $A_{12}$, then we multiply the equation (4-27)
from left side in $O_2^T$ and from the right hand in $O_2$, then
by comparing the two side of relation, we see that the matrix
$A_{22}$ can be diagonal by orthogonal matrix $O_2$ as
$$A_{22}=O_2D_{22}O_2^T$$
The similar result is obtained from equation (4-25) for the matrix
$A_{11}$:
$$A_{11}=O_1D_{11}O_1^T$$
By using the above result, the following transformation
for adjacency matrix is obtained
$$\left(\begin{array}{ccc}
          1& 0 & 0\\
            0& O_1^T& 0\\
            0 & 0 & O_2^T\\
          \end{array}\right)\left(\begin{array}{ccc}
         0& e_\kappa^T &0 \\
            e_\kappa& A_{11} & A_{12}\\
            0 & A_{12}^T & A_{22}\\
          \end{array}\right)\left(\begin{array}{ccc}
          1& 0 & 0\\
            0& O_1& 0\\
            0 & 0 & O_2\\
          \end{array}\right)=$$
\begin{equation}
          \left(\begin{array}{ccc}
          0& e_\kappa^TO_1 & 0\\
            O_1^Te_\kappa& O_1^TA_{11}O_1& O_1^TA_{12}O_2\\
            0 & O_2^TA_{12}^TO_1 & O_2^TA_{22}O_2\\
          \end{array}\right)\equiv \left(\begin{array}{ccccccc}
          0& \begin{array}{cccc}\sqrt{k}&0& \ldots &0\\ \end{array} & \begin{array}{ccc}0& \ldots &0\\ \end{array}\\
           \begin{array}{c}\sqrt{\kappa}\\ 0\\ \vdots \\0\\ \end{array} & D_{11} & D_{12}\\
            \begin{array}{c}0\\ \vdots \\0\\ \end{array} & D_{12}^T & D_{22}\\
          \end{array}\right)=D
\end{equation}
\textit{Case I: First strata} In this section we want to calculate
the adjacency matrix in the basis of first strata, So the equations (4-25),(4-27) and (4-28) should be rewritten for the nonzero
eigenvalue of matrix $J$, so
$$\sqrt{\lambda_{12}}(\lambda_1+\lambda_2)=\mu
\sqrt{\kappa(n-\kappa-1)}+(\lambda-\mu)\sqrt{\lambda_{12}}$$
$$\lambda_{12}+\lambda_2^2=(\kappa-\mu)+\mu(n-\kappa-1)+(\lambda-\mu)\lambda_2$$
\begin{equation}
\lambda_{12}+\lambda_1^2=(\kappa-\mu)+(\mu-1)k+(\lambda-\mu)\lambda_1
\end{equation}

Now  the $D^2$ from equation (4-29) must be calculate :
\begin{equation}
          D^2=\left(\begin{array}{ccccccc}
          k& (\begin{array}{cccc}\sqrt{\kappa}&0& \ldots &0\\ \end{array})D_{11} & (\begin{array}{cccc}\sqrt{\kappa}&0& \ldots &0\\ \end{array})D_{12}\\
           D_{11}\left(\begin{array}{c}\sqrt{\kappa}\\ 0\\ \vdots \\0\\ \end{array}\right) & \left(\begin{array}{cccc}\kappa&0& \ldots &0\\0&0&\ldots &0\\ \vdots &\vdots &\ddots &\vdots\\ 0&0&\ldots &0\\ \end{array}\right)+D_{11}^2+D_{12}D_{12}^T & D_{11}D_{12}+D_{12}D_{22}\\
            D_{12}^T\left(\begin{array}{c}\sqrt{\kappa}\\0\\ \vdots \\0\\ \end{array}\right) & D_{12}^TD_{11}+D_{22}D_{12}^T & D_{12}^TD_{12}+D_{22}^2\\
          \end{array}\right)
\end{equation}
$$\equiv(\kappa-\mu)I+\mu J+(\lambda - \mu)D$$
Where the matrix $J$ is in the form
\begin{equation}
J=\left(\begin{array}{ccccccc}
          1& \begin{array}{cccc}\sqrt{\kappa}&0& \ldots &0\\ \end{array} & \begin{array}{cccc}\sqrt{n-\kappa-1}&0& \ldots &0\\ \end{array}\\
           \begin{array}{c}\sqrt{\kappa}\\ 0\\ \vdots \\0\\ \end{array} & \begin{array}{cccc}\kappa&0& \ldots &0\\
           0&0&\ldots & 0\\
           \vdots & \vdots &\ddots &\vdots\\
           0&0&\ldots&0 \end{array} & \begin{array}{cccc}\sqrt{\kappa}\sqrt{n-\kappa-1}&0& \ldots &0\\
           0&0&\ldots & 0\\
           \vdots & \vdots &\ddots &\vdots\\
           0&0&\ldots&0 \end{array}\\
            \begin{array}{c}\sqrt{n-\kappa-1}\\0\\ \vdots \\0\\ \end{array} &  \begin{array}{cccc}\sqrt{\kappa}\sqrt{n-\kappa-1}&0& \ldots &0\\
           0&0&\ldots & 0\\
           \vdots & \vdots &\ddots &\vdots\\
           0&0&\ldots&0 \end{array}& \begin{array}{cccc}n-\kappa-1&0& \ldots &0\\
           0&0&\ldots & 0\\
           \vdots & \vdots &\ddots &\vdots\\
           0&0&\ldots&0 \end{array}\\
          \end{array}\right)
\end{equation}
from the block (3,1) and (2,1)
$$\sqrt{\kappa}\sqrt{\lambda_{12}}=\mu \sqrt{n-k-1}$$
\begin{equation}
\sqrt{k}\lambda_1=\lambda \sqrt{\kappa}
\end{equation}
So by substituting these results into equation (4-30) calculate the
parameters $\lambda_1$, $\lambda_2$ and $\lambda_{12}$ as
$$\lambda_1=\lambda$$
$$\lambda_2=\kappa-\mu$$
\begin{equation}
\lambda_{12}=\frac{\mu^2(n-\kappa-1)}{\kappa}
\end{equation}
So the adjacency matrix in the basis of first strata is
\begin{equation}
\left(\begin{array}{ccc}
         0& \sqrt{\kappa} & 0\\
            \sqrt{\kappa}& \lambda & \frac{\mu\sqrt{n-\kappa-1}}{\sqrt{\kappa}}\\
            0 & \frac{\mu\sqrt{n-\kappa-1}}{\sqrt{\kappa}} & \kappa - \mu\\
          \end{array}\right)
\end{equation}
Now we want to investigate the bipartite entanglement entropy in
SRGs in the case that the vertices of first and second strata are
in the first subset and the other vertices are in the second subset.  The
potential matrix is
\begin{equation}
V=\left(%
\begin{array}{cc}
  \widetilde{V}_{11} & \widetilde{V}_{12} \\
   \widetilde{V}_{12}^{T}& \widetilde{V}_{22} \\
\end{array}%
\right)
\end{equation}
By using (4-35) we have

$$\widetilde{V}_{11}=\left(%
\begin{array}{cc}
  1+2g\kappa & -2g\sqrt{\kappa} \\
  -2g\sqrt{\kappa} & 1+2g(\kappa-\lambda) \\
\end{array}%
\right),\widetilde{V}_{12}=\left(%
\begin{array}{c}
  0\\ -2g\mu\frac{\sqrt{n-\kappa-1}}{\sqrt{\kappa}} \\
\end{array}%
\right),\widetilde{V}_{22}=1+2g\mu$$
After applying the schur
complement method, the schmidt number of this case is
\begin{equation}
d_{(12,3)}^{(1)}=\frac{2\mu\sqrt{n-\kappa-1}\sqrt{1+2g\kappa}g}{\sqrt{\kappa}\sqrt{1+2g\mu}\sqrt{(1+2g\kappa)(1+2g(\kappa-\lambda))-4g^2\kappa}}
\end{equation}
Now, we investigate the case that the vertex of first stratum is
in the first subset and the other vertices are in the second subset. In this case the entanglement entropy between two subsets is obtained from only the first $3\times 3$ block of adjacency matrix, which is given in equation (4-35).
So, the potential matrix is
 $$\widetilde{V}_{11}=1+2g\kappa,\widetilde{V}_{12}=\left(%
\begin{array}{cc}
  -2g\sqrt{\kappa} & 0 \\
\end{array}%
\right),\widetilde{V}_{22}=\left(%
\begin{array}{cc}
  1+2g(\kappa-\lambda) & -2g\mu\frac{\sqrt{n-\kappa-1}}{\sqrt{\kappa}} \\
  -2g\mu\frac{\sqrt{n-\kappa-1}}{\sqrt{\kappa}} & 1+2g\mu \\
\end{array}%
\right)$$ We have:
\begin{equation}
d_{(1,23)}=\frac{2\sqrt{\kappa}\sqrt{1+2g\mu}g}{\sqrt{1+2g\kappa}\sqrt{(1+2g\mu)(1+2g(\kappa-\lambda))-4g^2\mu(\kappa-\lambda-1)}}
\end{equation}
the last case is that the vertices  of second stratum is in the first
subset and the other vertices are in the second subset. So, the
potential matrix is
$$\widetilde{V}_{11}=1+2g(\kappa-\lambda),\widetilde{V}_{12}=\left(%
\begin{array}{cc}
  -2g\sqrt{\kappa} & -2g\mu\frac{\sqrt{n-\kappa-1}}{\sqrt{\kappa}} \\
\end{array}%
\right),\widetilde{V}_{22}=\left(%
\begin{array}{cc}
  1+2g\kappa & 0 \\
  0 & 1+2g\mu \\
\end{array}%
\right)$$ We have:
\begin{equation}
d_{13,2}^{(1)}=2g\sqrt{\frac{\kappa^{2}}{(1+2g\kappa)(1+2g(\kappa-\lambda))}+\frac{\mu(\kappa-\lambda-1)}{(1+2g\mu)(1+2g(\kappa-\lambda))}}
\end{equation}

\textit{Case II: other strata}

The equations (4-25),
(4-27) and (4-28) for the other zero eigenvalues of matrix $J$
become
$$\sqrt{\lambda_{12}}(\lambda_1+\lambda_2)=(\lambda-\mu)\sqrt{\lambda_{12}}$$
$$\lambda_{12}+\lambda_2^2=(\kappa-\mu)+(\lambda-\mu)\lambda_2$$
\begin{equation}
\lambda_{12}+\lambda_1^2=(\kappa-\mu)+(\lambda-\mu)\lambda_1
\end{equation}
The solution for these equations is
\begin{equation}\label{adj.}
    \left\{\begin{array}{c}
      \hspace{-2.3cm}\quad \quad \lambda_1+\lambda_2=\lambda-\mu \quad\quad \quad \quad \quad\lambda_{12}\neq 0   \\
      \lambda_{1,2}^2=(\kappa-\mu)+(\lambda-\mu)\lambda_{1,2} \quad \quad\quad\lambda_{12}=0  \\
    \end{array}\right.
\end{equation}
And, the $2\times2$ matrix for other strata is obtained By solving
eigenvalue equation
$$\left(%
\begin{array}{cc}
  \lambda_{1}-x & \sqrt{\lambda_{12}} \\
  \sqrt{\lambda_{12}} & \lambda_{1}-x \\
\end{array}%
\right)$$ The eigenvalues of SRGs are
$x_{1,2}=\frac{1}{2}(\lambda-\mu\pm\sqrt{(\lambda-\mu)^{2}+4(\mu-\kappa)})$. So,
by using this fact that the sum of two eigenvalues is $\kappa-\mu$ and
the multiply of two eigenvalues is $\lambda-\mu$, we have
\begin{equation}
\lambda_{12}-\lambda_{1}\lambda_{2}=\kappa-\mu\quad\quad \quad
\quad \quad\lambda_{12}\neq 0
\end{equation}
So if one have one of the $\lambda_1$, $\lambda_2$ or
$\lambda_{12}$ for other strata, one can calculate the two other
parameters from the equations (4-41), (4-42).
\\The Schmidt number of other strata can be defined as following
\begin{equation}
d_{(2,3)}^{i\neq1}=\frac{2g\sqrt{\lambda_{12}}}{\sqrt{1+2g(\lambda_{12}-\lambda_{1})}\sqrt{1+2g(\lambda_{12}-\lambda_{2})}}
\end{equation}

\subsection{Entanglement entropy in the large coupling limit} In this section, our
derivation is based on the entanglement entropy for large coupling
strength. By using (4-38) We can rewrite the $d_{(1,23)}$ as
following
$$d_{(1,23)}^{(1)}=\frac{1}{\sqrt{1+\frac{1}{2g\kappa}}\sqrt{\frac{1}{2g}+\kappa-\lambda-\frac{2g\mu^{2}(n-\kappa-1)}{\kappa(1+2g\mu)}}}$$
Where
$\frac{2g\mu^{2}(n-\kappa-1)}{\kappa(1+2g\mu)}\simeq\frac{\mu(n-\kappa-1)}{\kappa}(1-\frac{1}{2g\mu})$
\\Therefore, by using $(2-6)$ we have
$$d_{(1,23)}^{(1)}\simeq\frac{1}{\sqrt{1+\frac{1}{2g\kappa}}\sqrt{1+\frac{n-1}{2g\kappa}}}\simeq1-\frac{1}{2}\varepsilon$$
And
$\varepsilon=\frac{1}{2g\kappa}+\frac{n-1}{2g\kappa}=\frac{n}{2g\kappa}$
\\By definition $(3-19)$, we can write
\begin{equation}
\gamma=\frac{1}{\sqrt{1-(1-\frac{1}{2}\varepsilon)^{2}}}\simeq\frac{1}{\sqrt{\varepsilon}}=\sqrt{\frac{2g\kappa}{n}}
\end{equation}
By using the definition of entanglement entropy, we have
\begin{equation}
S(\rho)=\frac{\gamma}{2}(1+\frac{1}{\gamma})\log\frac{\gamma}{2}(1+\frac{1}{\gamma})-\frac{\gamma}{2}(1-\frac{1}{\gamma})\log\frac{\gamma}{2}(1-\frac{1}{\gamma})
\end{equation}
$$=\frac{1}{2}((\gamma+1)(\log\frac{\gamma}{2}+\frac{1}{\gamma}))-\frac{1}{2}((\gamma-1)(\log\frac{\gamma}{2}-\frac{1}{\gamma}))$$
So
\begin{equation}
S(\rho)_{1,23}=\log\frac{\gamma}{2}+1=\frac{1}{2}\log\frac{g\kappa}{2n}+1
\end{equation}
Where $\kappa$ is the size of the boundary between the first and the
second subsets. So, we see that the entanglement entropy has a logarithmic
relation with the ratio of size of boundary to the size of the system.
\\We can calculate the above result for (4-37) and (4-39). We can
rewrite the $d_{(12,3)}^{(1)}$ by using By using (4-37) and (2-6), as following form
$$d_{(12,3)}^{(1)}=\frac{1}{\sqrt{1+\frac{1}{2g\mu}}\sqrt{\frac{1}{2g(\kappa-\lambda-1)}+\frac{\kappa-\lambda}{\kappa-\lambda-1}\frac{2g\kappa}{(\kappa-\lambda-1)(1+2g\kappa)}}}$$
Where
$\frac{2g\kappa}{(\kappa-\lambda-1)(1+2g\kappa)}\simeq\frac{1}{(\kappa-\lambda-1)}(1-\frac{1}{2g\kappa})$
\\So
$$d_{(12,3)}^{(1)}\simeq\frac{1}{\sqrt{1+\frac{1}{2g\mu}}\sqrt{1+\frac{\kappa+1}{2g\mu(n-\kappa-1)}}}=1-\frac{1}{2}\varepsilon$$
And
$\varepsilon=\frac{1}{2g\mu}+\frac{\kappa+1}{2g\mu(n-\kappa-1)}=\frac{n}{2g\mu(n-\kappa-1)}$
\\By the same way and by $(4-45)$ and $(4-46)$, we have
\begin{equation}
S(\rho)_{12,3}=\log\frac{\gamma}{2}+1=\frac{1}{2}\log\frac{\mu
g(n-\kappa-1)}{2n}+1
\end{equation}
Where $\mu(n-\kappa-1)$ is the size of boundary.
\subsection{Examples: Some important kinds of SRG classes by using stratification which don't contain
nonisomorph SRGs}
In this section we want to study the entanglement entropy for some
kinds of SRGs. We could identify their adjacency matrices in the stratification basis.
Two graphs will be isomorphic, when those are related to each
other by a relabeling of vertices.
For these kinds of SRGs, there are not any nonisomorph SRG.
\\\textit{example I}:Normal subgroup graph $(2m,m,0,m)$
\\Let $G$ be a finite group, and $P=P_{0},P_{1},...,P_{d}$ be a
blueprint of it. we always assume that the sets $P_{i}$ are so
numbered that the identity element $e$ of $G$ belongs to $P_{0}$,
if $P_{0}=e$, the $P$ is called homogeneous. Let
$R_{0},R_{1},...,R_{d}$ be the set of relations
$R_{i}={(\alpha,\beta)\in G\otimes G|\alpha^{-1}\beta\in P_{i}}$
on $G$. Now, we define a blueprint for group $G$ which form a
strongly regular graph. If $H$ is a subgroup of $G$ , we define
the blueprints by
\begin{equation}
P_{0}=e,P_{1}=G-H,p_{2}=H-e
\end{equation}
This blueprint form a strongly regular graph with parameters
$(n,\kappa,\lambda,\mu)=(|G|,|G|-|H|,|G|-2|H|,|G|-|H|)$. As an
example, we consider $G=D_{2m}(m=odd)$
\begin{equation}
H={e,a,a^{-1},...,a^{(m-1)/2},a^{-(m-1)/2}}
\end{equation}
Therefore the blueprints are given by
\begin{equation}
P_{0}={e},P_{1}={b,ab,a^{2}b,...,a_{m-1}b},P_{2}={e,a,a^{-1},...,a^{(m-1)/2},a^{-(m-1)/2}}
\end{equation}
Which form a strongly regular graphs with parameters $(2m,m,0,m)$.
The stratification basis for this graph are
$$|\phi_{0}\rangle=|e\rangle$$
$$|\phi_{1}\rangle=\frac{1}{\sqrt{m}}\sum_{i=0}^{m-1}|a^{i}b\rangle$$
\begin{equation}
|\phi_{2}\rangle=\frac{1}{\sqrt{m-1}}\sum_{i=0}^{m-1}|a^{i}\rangle
\end{equation}
The action of adjacency matrix on the stratification basis is
$$A|\phi_{0}\rangle=\sqrt{m}|\phi_{1}\rangle$$
$$A|\phi_{1}\rangle=\sqrt{m}|\phi_{0}\rangle+\sqrt{m(m-1)}|\phi_{2}\rangle$$
\begin{equation}
A|\phi_{2}\rangle=\sqrt{m(m-1)}|\phi_{1}\rangle
\end{equation}
So, the adjacency matrix is
\begin{equation}
\left(%
\begin{array}{ccc}
  0 & \sqrt{m} & 0 \\
  \sqrt{m} & 0 & \sqrt{m(m-1)} \\
  0 & \sqrt{m(m-1)} & 0 \\
\end{array}%
\right)
\end{equation}
After the generalized Schur complement method,
the schmidt numbers are
\begin{equation}
d_{1,23}^{(1)}=\frac{2\sqrt{m}g}{\sqrt{(1+2gm)^{2}-4m(m-1)g^{2}}}
\end{equation}
And
\begin{equation}
d_{12,3}^{(1)}=\frac{2\sqrt{m(m-1)}g}{\sqrt{(1+2gm)^{2}-4mg^{2}}}
\end{equation}
And
\begin{equation}
d_{13,2}^{(1)}=\frac{2g}{1+2gm}\sqrt{m(2m-1)}
\end{equation}
\\\textit{example II}: $\kappa = \mu$  $:(2k-\lambda, k, \lambda, k)$
\\For the case $\kappa=\mu$
from equation (4-26),
$$A_{22}=0$$ So from (4-27), $$A_{12}^TA_{12}=\kappa J_{k'k'}$$
So, $A_{12}=J_{\kappa\acute{\kappa}}$.\\ The second
equation for the diagonal elements is rewritten:
\begin{equation}
\lambda+\acute{\kappa}-\mu+1=0
\end{equation}
By substituting $n-\kappa-1$ for $\acute{\kappa}$
\begin{equation}
n=2\kappa-\lambda
\end{equation}
It can be shown that by considering the other elements of second
equation, again  the same equation as diagonal
elements$(n=2\kappa-\lambda)$ are obtained.\\So, the parameters is
$(2\kappa-\lambda,\kappa,\lambda,\kappa)$. \\ Therefore we
conclude that the Schmidt number and entanglement entropy is
obtained from the first stratum. It's clear that the entanglement
entropy can not distinguish two non-isomorphic graphs of these
kinds. the parameter $d_{(12,3)}^{(1)}$ is obtained just from
(4-37) by substituting $\kappa$ instead of $\mu$
\begin{equation}
d_{(12,3)}^{(1)}=\frac{2\sqrt{n-\kappa-1}\sqrt{\kappa}g}{\sqrt{(1+2g\kappa)(1+2g(k-\lambda))-4g^2k}}
\end{equation}
\textit{example III}: $\lambda=0 : (\frac{k(k-1)}{\mu}+k+1, k, 0, \mu)$
 \\In this case, from equation
(4-24) we find that
$$A_{11}=0$$
So by substituting $\lambda=0$ in the third equations of (4-30),
the eigenvalues of $A_{12}$ for the first stratum and other strata
are
$$\lambda_{12}^{(1)}=\mu(\kappa-1)$$
\begin{equation}
\lambda_{12}^{(i\neq 1)}=\kappa-\mu
\end{equation}
We have explained the case $\kappa=\mu$ in the previous example,
So we suppose that $\kappa\neq \mu$, therefore $\lambda_{12} \neq
0$. So from (4-41) we find that $$\lambda_2=\lambda-\mu=-\mu$$ In
this case also the Schmidt number is related to parameters of SRG,
So it can not distinguish non-isomorph graphs the same as the
previous example.\\The parameter $d_{(2,3)}^{(i \neq 1)}$ for
these kind of graphs become
\begin{equation}
d_{(2,3)}^{(i \neq
1)}=\frac{2g\sqrt{\kappa-\mu}}{\sqrt{1+2g\kappa}\sqrt{1+2g(\kappa-\mu)}}
\end{equation}
\textit{example IV}: $A_{12}A_{12}^{T}=\kappa$
\\Now we want to investigate the SRG graphs these kind which their
$A_{12}$ is $\kappa\times1$ complete graph. So
\begin{equation}
A_{12}^{T}A_{12}=\kappa
\end{equation}
 And this case satisfy the equations (4-25), (4-27) and (4-28) by considering
 $$\lambda=\kappa-2$$
\begin{equation}
n=\kappa+2
\end{equation}
Therefore the parameters of this case will be:
$$(\kappa+2,\kappa,\kappa-2,\kappa)$$
We know that it is possible to write the matrices $A_11$
and $A_22$ in terms of the matrix
representations of permutation group, So suppose
$$A_{11}=J-I-\pi$$
$$A_{11}^{2}=(\kappa-4)J+I+\pi^{2}+2\pi$$
Also from(4-28), we find that
$$A_{11}^{2}=(\kappa-2)J-2A_{11}$$
After comparing the two above equation
$$\pi^{2}=I$$
We conclude that $\pi$ is an element of cycle group with order two, therefore the parameter $\kappa$
can not be odd.
By substituting the parameters of these kinds of SRG into (2-8) we find
$$A^{2}=\kappa J-2A$$
By comparing this relation for $A^{2}$
with relation for $A_{11}^{2}$
, and one conclude that the matrix $A_{11}$ of
this graph for the case with degree $\kappa$, is the Adjacency matrix of these kinds of graphs with
degree $\kappa-2$.
Therefore the third stratum, contains only one vertex, So these kinds of graphs can not
distinguish the nonisomorph graphs.
\subsection{examples:Some important kinds of SRGs which contain nonisomorph SRGs}
For some important kinds of SRGs, we could identified their adjacency matrices in the stratification basis. Then we investigate the graph isomorphism problem by using the blocks of
adjacency matrices in the stratification basis analytically.
\\\textit{example I}: Triangular graph
$(\frac{\nu(\nu-1)}{2},2(\nu-2),\nu-2,4)$
\\for positive integer $\nu$ the triangular graph $T_{n}$ is
strongly regular graph.As the construction is completely
symmetric, we may begin by considering any vertex, say the one
labeled by the set $(1,2)$.Every vertex labeled by a set of form
$(1,i)$ or $(2,i)$, for $i\geq3$, will be connected to this
set.So, this vertex, and every vertex, has degree $2(\nu-2)$. For
any neighbor of $(1,2)$,say $(1,3)$, every other vertex of form
$(1,i)$ for $i\geq4$ will be a neighbor of both of these, as will
the set of $(2,3)$. Carrying this out in general, we find that
$\lambda=\nu-2$. Finally any non-neighbor of $(1,2)$, say $(3,4)$,
will have $4$ common neighbors with $(1,2)$ [13]. So, $\mu=4$ and $n=\left(%
\begin{array}{c}
  \nu \\
  2 \\
\end{array}%
\right)$ \\In triangular graph the $A_{11}$ is defined as
following form:
\begin{equation}
A_{11}=I_{2}\otimes(J_{\nu-2}-I_{\nu-2})+X\otimes I_{\nu-2}
\end{equation}
where $X=\left(%
\begin{array}{cc}
  0 & 1 \\
  1 & 0 \\
\end{array}%
\right)$.\\And the eigenvalues of $A_{11}$ is
$$\lambda_{1}={\nu-2,\nu-4,\overbrace{0}^{\nu-3},\overbrace{-2}^{\nu-3}}$$
$\nu-2$ is the biggest eigenvalue. By using it, we can calculate the $3 \times 3$ block of adjacency matrix as
\begin{equation}
\left(%
\begin{array}{ccc}
  0 & \sqrt{2(\nu-2)} & 0 \\
  \sqrt{2(\nu-2)} & \nu-2 & 2\sqrt{(\nu-3)} \\
  0 & 2\sqrt{(\nu-3)} & 2(\nu-4) \\
\end{array}%
\right)
\end{equation}
The schmidt number between each two parts in first strata
obtained from equation (4-37),(4-38) and(4-39)
\begin{equation}
d_{1,23}^{(1)}=\frac{2g\sqrt{2(\nu-2)(1+8g)}}{\sqrt{1+4g(\nu-2)}\sqrt{(1+8g)(1+2g(\nu-2))-16g^{2}(\nu-3)}}
\end{equation}
 And
\begin{equation}
d_{12,3}^{(1)}=\frac{4g\sqrt{(\nu-3)(1+4g(\nu-2))}}{\sqrt{1+8g}\sqrt{(1+4g(\nu-2))(1+2g(\nu-2))-8g^{2}(\nu-3)}}
\end{equation}
and
\begin{equation}
d_{13,2}^{(1)}=4g\sqrt{\frac{(\nu-2)^{2}}{(1+4g(\nu-2))(1+2g(\nu-2))}+\frac{\nu-3}{(1+8g)(1+2g(\nu-2))}}
\end{equation}
By using other eigenvalues and the
equation (4-42), we see that the eigenvalues $\nu-4,-2$ are
singlet. So, we can calculate other strata of triangular graph by
$\lambda_{1}=0$.
\begin{equation}
\left(%
\begin{array}{cc}
  0 & \sqrt{2(\nu-4)} \\
  \sqrt{2(\nu-4)} & \nu-6 \\
\end{array}%
\right)
\end{equation}
By using the Schur complement method, the schmidt number can be calculated as
following
\begin{equation}
d_{(2,3)}^{(2)}=\frac{2g\sqrt{2(\nu-4)}}{\sqrt{1+4g(\nu-4)}\sqrt{1+2g(\nu-2)}}
\end{equation}
And the entanglement entropy can be obtained from equation (3-18)
and (3-19). \\ The strong regular graph with parameters
$(28,12,6,4)$ have $4$ non-isomorphic graphs that one of them is
triangular graph.
\\\textit{example II}: Lattice graphs $(\nu^{2},2(\nu-1),\nu-2,2)$
\\For positive integer $\nu$, the lattice graph $L_{n}$ is the
graph with vertex set ${1,...,\nu}^{2}$ in which vertex $(a,b)$ is
connected to vertex $(c,d)$ if $a=c$ or $b=d$.Thus the vertices
may be arranged at the points in an $\nu-$by$-\nu$ grid, with
vertices being connected if they lie in the same row or column. It
is routine to see that the parameters of this graph are:
\begin{equation}
\kappa=2(\nu-1),\lambda=\nu-2,\mu=2
\end{equation}
In lattice graph adjacency matrix is
\begin{equation}
A=I_{\nu}\otimes(J_{\nu}-I_{\nu})+(J_{\nu}-I_{\nu})\otimes I_{\nu}
\end{equation}
So, the $A_{11}$ is defined as following form:
\begin{equation}
A_{11}=I_{2}\otimes(J_{\nu-1}-I_{\nu-1})
\end{equation}
And the eigenvalues of $A_{11}$ are
$$\lambda_{1}={\overbrace{\nu-2}^{2},\overbrace{-1}^{2(\nu-2)}}$$
$\nu-2$ is the biggest eigenvalue. By using it, we can calculate the $3 \times 3$ block of adjacency matrix as
\begin{equation}
\left(%
\begin{array}{ccc}
  0 & \sqrt{2(\nu-1)} & 0 \\
  \sqrt{2(\nu-1)} & \nu-2 & \sqrt{2(\nu-1)} \\
  0 & \sqrt{2(\nu-1)} & 2(\nu-2) \\
\end{array}%
\right)
\end{equation}
As example I the entanglement entropy between each two parts in the
first strata obtained from equation (4-37),(4-38)and (4-39).
\begin{equation}
d_{1,23}^{(1)}=\frac{2g\sqrt{2(\nu-1)(1+4g)}}{\sqrt{1+4g(\nu-1)}\sqrt{(1+4g)(1+2g\nu)-8g^{2}(\nu-1)}}
\end{equation}
 And
\begin{equation}
d_{12,3}^{(1)}=\frac{2g\sqrt{2(\nu-1)(1+4g(\nu-1))}}{\sqrt{1+4g}\sqrt{(1+4g(\nu-1))(1+2g\nu)-8g^{2}(\nu-1)}}
\end{equation}
and
\begin{equation}
d_{13,2}^{(1)}=2g\sqrt{\frac{4(\nu-1)^{2}}{(1+4g(\nu-1))(1+2g\nu)}+\frac{2(\nu-1)}{(1+4g)(1+2g\nu)}}
\end{equation}
\\By using other eigenvalues and the equation $(4-42)$, we can
calculate other strata of lattice graph. By $\lambda_{1}=-1$ the
second strata is
\begin{equation}
\left(%
\begin{array}{cc}
  -1 & \sqrt{(\nu-1)} \\
  \sqrt{(\nu-1)} & \nu-3 \\
\end{array}%
\right)
\end{equation}
The schmidt number can be calculated as
following
\begin{equation}
d_{(2,3)}^{(2)}=\frac{2g\sqrt{(\nu-1)}}{\sqrt{1+4g(\nu-1)}\sqrt{1+2g(\nu-1)}}
\end{equation}
And the entanglement entropy can be obtained from equation (3-18)
and (3-19).\\ The strong regular graph with parameters
$(16,6,2,2)$ have $2$ non-isomorphic graphs that one of them is
lattice graph.
\\\textit{example III}: Latin Square graphs $(\nu^{2},3(\nu-1),\nu,6)$
\\A Latin Square is an $\nu-$by$-\nu$ grid, each entry of which is
a number between $1$ and $\nu$, such that no number appears twice
in any row or column. So, it will have $\nu^{2}$ nodes, one for each
cell in the square. Two nodes are joined by an edge if
\\$1$. they are in the same row,
\\$2$. they are in the same column, or
\\$3$. they hold the same number.
\\So, such a graph has degree $\kappa=3(\nu-1)$. Any two nodes in
the same row will both be neighbors with every other pair of nodes
in their row. They will have two more common neighbors: The nodes
in their columns holding the other's number. So, they have $\nu$
common neighbors. The same obviously holds for columns, and is
easy to see for nodes that have the same number.So, every pair of
nodes that are neighbors have exactly $\lambda=\nu$ common
neighbors. On the other hand, consider two vertices that are not
neighbors,they lie in different rows, lie in different columns,
and hold different numbers. So, $\mu=6$. \\In latin square graph
adjacency matrix is
\begin{equation}
A=I_{\nu}\otimes(J_{\nu}-I_{\nu})+(J_{\nu}-I_{\nu})\otimes
I_{\nu}+\sum_{k=1}^{\nu}S^{k}\otimes S^{n-k}
\end{equation}
Where $S$ is shift operator. So, the $A_{11}$ is defined as
following form:
\begin{equation}
A_{11}=I_{3}\otimes(J_{\nu-1}-I_{\nu-1})+(J_{3}-I_{3})\otimes
f_{\nu-1}
\end{equation}
Where $f_{\nu}=\left(%
\begin{array}{cccc}
  0 & 0 & ... & 1 \\
  0 & ... & 1 & 0 \\
  \vdots & \ddots & \vdots & \vdots \\
  1 & 0 & ... & 0 \\
\end{array}%
\right)$ is off-diagonal matrix.
\\If $\nu=2l(\nu$ is even), the eigenvalues of $A_{11}$ is
$$\lambda_{1}={\nu,\overbrace{\nu-3}^{2},\overbrace{1}^{l-1},\overbrace{0}^{(\nu-2)},\overbrace{-2}^{(\nu-2)},\overbrace{-3}^{l-1}}$$
$\nu$ is the biggest eigenvalue. So the $3 \times 3$ block of adjacency matrix is
\begin{equation}
\left(%
\begin{array}{ccc}
  0 & \sqrt{3(\nu-1)} & 0 \\
  \sqrt{3(\nu-1)} & \nu & 2\sqrt{3(\nu-2)} \\
  0 & 2\sqrt{3(\nu-2)} & 3(\nu-3) \\
\end{array}%
\right)
\end{equation}
Again, the entanglement entropy between each two parts
in the first stratum, is obtained from equation (4-37),(4-38) and (4-39). So we
have
\begin{equation}
d_{1,23}^{(1)}=\frac{2g\sqrt{3(\nu-1)(1+12g)}}{\sqrt{1+6g(\nu-1)}\sqrt{(1+12g)(1+2g(2\nu-3))-48g^{2}(\nu-2)}}
\end{equation}
 And
\begin{equation}
d_{12,3}^{(1)}=\frac{4g\sqrt{3(\nu-2)(1+6g(\nu-1))}}{\sqrt{1+12g}\sqrt{(1+6g(\nu-1))(1+2g(2\nu-3))-12g^{2}(\nu-1)}}
\end{equation}
and
\begin{equation}
d_{13,2}^{(1)}=2g\sqrt{\frac{9(\nu-1)^{2}}{(1+6g(\nu-1))(1+2g(2\nu-3))}+\frac{12(\nu-2)}{(1+12g)(1+2g(2\nu-3))}}
\end{equation}
 By using other eigenvalues and the equation (4-42), we see that $\lambda_{1}=\nu-3,-3$ are singlets. So, we can
calculate other strata of latin square graph by
$\lambda_{1}=1,0,-2$. the other strata are
\begin{equation}
\left(%
\begin{array}{cc}
  1 & \sqrt{4(\nu-4)} \\
  \sqrt{4(\nu-4)} & \nu-7 \\
\end{array}%
\right)
\end{equation}
and
\begin{equation}
\left(%
\begin{array}{cc}
  0 & \sqrt{3(\nu-3)} \\
  \sqrt{3(\nu-3)} & \nu-6 \\
\end{array}%
\right)
\end{equation}
and
\begin{equation}
\left(%
\begin{array}{cc}
  -2 & \sqrt{(\nu-1)} \\
  \sqrt{(\nu-1)} & \nu-4 \\
\end{array}%
\right)
\end{equation}
So the parameters $d_{(2,3)}^{(i)}$ where $i=2,3,\ldots,l+\nu-1$ are
\begin{equation}
d_{(2,3)}^{(2)}=\frac{2g\sqrt{4(\nu-4)}}{\sqrt{1+2g(4\nu-17)}\sqrt{1+6g(\nu-3)}}
\end{equation}
\begin{equation}
d_{(2,3)}^{(l+1)}=\frac{2g\sqrt{3(\nu-3)}}{\sqrt{1+2g(2\nu-3)}\sqrt{1+6g(\nu-3)}}
\end{equation}
\begin{equation}
d_{(2,3)}^{(l+\nu-1)}=\frac{2g\sqrt{(\nu-1)}}{\sqrt{1+2g(\nu+1)}\sqrt{1+6g}}
\end{equation}
And the entanglement entropy can be obtained from equation (3-18)
and (3-19).
\\If $\nu=2l+1(\nu$ is odd), the eigenvalues of $A_{11}$ are
$$\lambda_{1}={\nu,\overbrace{\nu-3}^{2},\overbrace{1}^{l-1},\overbrace{-2}^{(\nu-3)},\overbrace{0}^{(\nu-1)},\overbrace{-3}^{l}}$$
The strata of this case ($\nu$ is odd) is the same as first case ($\nu$ is even).
\\ The strong regular graph with parameters
$(16,9,4,6)$ have $2$ non-isomorphic graphs and $(25,12,5,6)$ have
$15$ non-isomorphic graphs and $(49,18,7,6)$ have $147$
non-isomorphic graphs that one of them is latin square graph.
\\\textit{example IV}: Generalized
Quadrangle $GQ(s,t)$, $((st+1)(s+1),s(t+1),s-1,t+1)$
\\A Generalized Quadrangle $GQ(s,t)$ is an incidence structure of
points and lines with the following properties $[21]$
\\$1$. Every line has $s+1$ points and every point is on $t+1$
lines.
\\$2$. Any two distinct points are incident with at most one line.
\\$3$. Given a line $L$ and a point $p$ not on $L$, there is a
unique point on $L$ collinear with $p$(two points are said to be
collinear if there is a line incident with both).
\\Its strongly regular graph 's parameter set is
$((st+1)(s+1),s(t+1),s-1,t+1)$. Necessary conditions for existence
of a $GQ(s,t)$ are $1\leq t\leq s^{2}$ if $s>1$, and $s+t$ divides
$st(s+1)(t+1)$. So,the $A_{11}$ is defined as following form:
\begin{equation}
A_{11}=I_{t+1}\otimes(J_{s}-I_{s})
\end{equation}
$$\lambda_{1}={\overbrace{s-1}^{t+1},\overbrace{-1}^{(t+1)(s-1)}}$$
$s-1$ is the biggest eigenvalue. So the $3 \times 3$ block of adjacency matrix is
\begin{equation}
\left(%
\begin{array}{ccc}
  0 & \sqrt{s(t+1)} & 0 \\
  \sqrt{s(t+1)} & s-1 & \sqrt{st(t+1)} \\
  0 & \sqrt{st(t+1)} & (s-1)(t+1) \\
\end{array}%
\right)
\end{equation}
We have
\begin{equation}
d_{1,23}^{(1)}=\frac{2g\sqrt{s(t+1)(1+2g(t+1)}}{\sqrt{1+2gs(t+1)}\sqrt{(1+2g(t+1))(1+2g(st+1))-4g^{2}st(t+1)}}
\end{equation}
And
\begin{equation}
d_{12,3}^{(1)}=\frac{2g\sqrt{st(1+t)(1+2gs(t+1))}}{\sqrt{1+2g(t+1)}\sqrt{(1+2gs(t+1))(1+2g(st+1))-4g^{2}s(t+1)}}
\end{equation}
and
\begin{equation}
d_{13,2}^{(1)}=2g\sqrt{\frac{s^{2}(t+1)^{2}}{(1+2gs(t+1))(1+2g(st+1))}+\frac{st(t+1)}{(1+2g(t+1))(1+2g(st+1))}}
\end{equation}
\\By using other eigenvalues and the equation (4-42), we
can calculate other strata of generalized quadrangle graph. The
second strata is
\begin{equation}
\left(%
\begin{array}{cc}
-1 & \sqrt{st} \\
  \sqrt{st} & s-t-1 \\
\end{array}%
\right)
\end{equation}
The parameter $d_{(2,3)}^{(2)}$ is
\begin{equation}
d_{(2,3)}^{(2)}=\frac{2g\sqrt{st}}{\sqrt{1+2g(1+st)}\sqrt{1+2g(s(t-1)+t+1)}}
\end{equation}
And the entanglement entropy can be obtained from equation (3-18)
and (3-19).
\\The strong regular graph with parameters $(40,12,2,4)$ have $28$ non-isomorphic graphs and $(45,12,3,3)$ have
$78$ non-isomorphic graphs and $(64,18,2,6)$ have 167
non-isomorphic graphs that one of them is generalized quadrangle
graph.
\subsection{Area-Law}
Entanglement entropy is a quantitative measure of the quantum entanglement. A natural problem then, is to divide the system into two
regions and study the entanglement entropy between them. In general, short-range correlations, which are
non-universal, give a contribution proportional to the area of the boundary between the two regions to the
entanglement entropy. This is often called as the area-law contribution. In one dimension, the area-law contribution
is constant with respect to the system size or to the
size of the regions.  For a regular
lattice, the size of the boundary of an element is given by
twice its dimensionality thus, in analogy, for a node in a
complex network its boundary  is given by
its connectivity.
\\In our system, the area law is studied in bipartite systems. Two case will be choose
\\\textit{Case I}: $\mu$ is finite and $\lambda, \kappa$ are infinite.
\\When $\lambda,\kappa$ are infinite, it means that the size of the system is infinite. The parameter $\gamma$ from (3-19) can be written as
\begin{equation}
\gamma_{(1,23)}^{(1)}=\sqrt{\frac{1+2g\kappa}{1+2g\kappa-\frac{4g^{2}\kappa(1+2g\mu)}{1+4g^{2}\mu+2g\mu(\kappa-\lambda+\mu)}}}
\end{equation}
By finite $\mu$, in highest connectivity, the statement $\frac{4g^{2}\kappa(1+2g\mu)}{1+4g^{2}\mu+2g\mu(\kappa-\lambda+\mu)}$ tended
to zero  and parameter $\gamma$ tended to one. So, in the large size of the system, we don't have large amount of entanglement entropy.
\\For example, in triangle graph (4-66), lattice graph (4-75) and latin square graph(4-83),
for infinite $\nu$, it is clear that the
schmidt number $d_{(1,23)}^{(1)}\rightarrow 0$. So the parameter $\gamma$ tends to one.
entanglement entropy $S(\rho)_{(1,23)}\rightarrow 0$. So, there is no entanglement between
strata.
\\\textit{Case II}: $\lambda$ is finite and $\kappa=\mu$ is infinite
\\In this case, the parameter $\gamma$ from (3-18) can be written as
\begin{equation}
\gamma_{(1,23)}^{(1)}=\sqrt{\frac{1+2g\kappa}{1+2g\kappa-\frac{4g^{2}\kappa(1+2g\kappa)}{1+4g^{2}\kappa+2g\kappa(2\kappa-\lambda)}}}
\end{equation}
Also, in this case the parameter $\gamma$ tends to one and the
entanglement entropy $S(\rho)_{(1,23)}\rightarrow 0$. So, there is no entanglement between
strata.

\section{Investigation of graph isomorphism problem in SRGs}
Two graph will be isomorphic, when those are related to each other
by a relabeling of vertices.two non isomorphic graphs could be
distinguished with quantum random  walk in ref [13]. Here, we want
to investigate the graph isomorphism problem by using different
eigenvalues of the matrix $A_{12}$. Our method can distinguish
non-isomorphic graphs with simple method.
\\There are some non-isomorphic graphs with SRG parameters, which
their $\lambda_{12}$s are different.
$$(n,\kappa,\lambda,\mu)=(25,12,5,6)$$.
There are $6$ different eigenvalues of the matrix $A_{12}$.
$$\lambda_{12}(1)=6,\overbrace{2.4495}^{3},2.3268,2.1753,2,1.6080,1.1260,\overbrace{0}^{3}$$
$$\lambda_{12}(2)=6,\overbrace{2.4495}^{4},\overbrace{1.7321}^{4},\overbrace{0}^{3}$$
$$\lambda_{12}(3)=6,\overbrace{2.4495}^{4},\overbrace{2.1753}^{2},\overbrace{1.1260}^{2},\overbrace{0}^{3}$$
$$\lambda_{12}(4)=6,\overbrace{2.4495}^{2},\overbrace{2.2770}^{2},\overbrace{2}^{3},\overbrace{0.7672}^{2},\overbrace{0}^{2}$$
$$\lambda_{12}(5)=6,\overbrace{2.4495}^{4},\overbrace{2}^{3},\overbrace{0}^{4}$$
$$\lambda_{12}(6)=6,\overbrace{2}^{9},\overbrace{0}^{2}$$
Another graph is
$$(n,\kappa,\lambda,\mu)=(26,10,3,4)$$.
There are $5$ different eigenvalues of the matrix $A_{12}$.
$$\lambda_{12}(1)=4.8990,2.4972,2.3073,\overbrace{2.2361}^{4},1.3556,1.3281,0.5645$$
$$\lambda_{12}(2)=4.8990,\overbrace{2.4495}^{2},\overbrace{2}^{6},0$$
$$\lambda_{12}(3)=4.8990,\overbrace{2.4994}^{2},2.4812,\overbrace{2.1342}^{2},\overbrace{1.7883}^{2},1.1701,0.6889$$
$$\lambda_{12}(4)=4.8990,\overbrace{2.4953}^{2},\overbrace{2.2770}^{2},\overbrace{2}^{3},\overbrace{0.7672}^{2}$$
$$\lambda_{12}(5)=4.8990,\overbrace{2.4495}^{4},\overbrace{2}^{3},\overbrace{0}^{2}$$
Another graph is
$$(n,\kappa,\lambda,\mu)=(28,12,6,4)$$.
There are $4$ different eigenvalues of the matrix $A_{12}$.
$$\lambda_{12}(1)=\sqrt{20},\overbrace{2.9356}^{2},\overbrace{2.5263}^{2},\overbrace{2.2361}^{2},\overbrace{0}^{5}$$
$$\lambda_{12}(2)=\sqrt{20},\overbrace{\sqrt{8}}^{5},\overbrace{0}^{6}$$
$$\lambda_{12}(3)=\sqrt{20},\overbrace{\sqrt{8}}^{4},\overbrace{2}^{2},\overbrace{0}^{5}$$
$$\lambda_{12}(4)=\sqrt{20},\overbrace{\sqrt{8}}^{2},\overbrace{2.4495}^{4},\overbrace{0}^{5}$$
Another graph is
$$(n,\kappa,\lambda,\mu)=(36,14,4,6)$$.
There are $3$ different eigenvalues of the matrix $A_{12}$.
$$\lambda_{12}(1)=7.3485,2.9849,2,9832,2.9713,2,9244,2.8810,2.7777,2.6722,2.2143$$
$$,2.1213,1.7809,0.7695,0.6985,0$$
$$\lambda_{12}(2)=7.3485,\overbrace{2.9713}^{4},\overbrace{2.8284}^{3},\overbrace{1.7809}^{4},\overbrace{0}^{2}$$
$$\lambda_{12}(3)= 7.3485,2.9863,2.9785,2.9356,2.9173,2.7501,2.5354,2.5263,2.3189,2.2998$$
$$2.0165,1.2072,0.7204,0$$
Another graph is
$$(n,\kappa,\lambda,\mu)=(40,12,2,4)$$.
There are $4$ different eigenvalues of the matrix $A_{12}$.
$$\lambda_{12}(1)= 6,\overbrace{3}^{4},\overbrace{2.8284}^{2},\overbrace{2.2361}^{4},0$$
$$\lambda_{12}(2)= 6,\overbrace{3}^{6},2.8284,\overbrace{2.2361}^{2},\overbrace{0}^{2}$$
$$\lambda_{12}(4)=6,\overbrace{3}^{8},\overbrace{0}^{3}$$
Another graph is
$$(n,\kappa,\lambda,\mu)=(50,21,8,9)$$.
There are $10$ different eigenvalues of the matrix $A_{12}$.
$$\lambda_{12}(1)=10.3923,3.4971,3.4681,3,4568,3.4514,3.3775,3.2582,3.1566$$
$$,3.1279,2.9672,2.9551,2.8718,2.6479,2.4173,2.1405,1.9254,1.7811,1.2576,1.2507$$
$$0.9902,0.1757$$
$$\lambda_{12}(2)=10.3923,3.4998,3.4873,3.4097,3.4095,3.3335,3.3019,3.2827$$
$$,3.2455,2.9175,2.8205,2.7827,2.6663,2.3496,2.1026,1.9616,1.7264,1.6747,1.1559$$
$$0.8272,0.2970$$
$$\lambda_{12}(3)=10.3923,3.5,3.4826,3.4790,3.4612,3.3755,3.2339,3.1841$$
$$3.1585,3.0605,3.0311,2.5686,2.5446,2.4968,2.3267,2.2466,1.3636,1.3571$$
$$0.9672,0.7667,0.1583$$
$$\lambda_{12}(4)=10.3923,3.4963,3.4945,3.4877,3.47,3.4433,3.2566,3.1966$$
    $$3.1858,2.9511,2.9340,2.5904,2.5879,2.3506,1.9505,1.9489,1.8770,1.7911$$
    $$1.1268,0.8170,0.0947$$
$$\lambda_{12}(5)= 10.3923,3.4757,3.4589,3.4482,3.3634,3.3024,3.2533,3.1747$$
    $$3.0016,2.9364,2.7631,2.7481,2.6822,2.6359,2.4051,1.9295,1.8550,1.7033$$
$$1.0740,1.0527,0.3333$$
$$\lambda_{12}(6)= 10.3923,3.4978,3.4826,3.4741,3.3731,3.2942,3.2315,3.2165$$
    $$3.1663,2.9801,2.9188,2.5423,2.5208,2.2940,2.2563,2.0890,1.8087,1.7151$$
   $$ 1.1861,1.1438,0.8130$$
$$\lambda_{12}(7)= 10.3923,4.0749,3.4989,3.4975,3.4216,3.3640,3.2608,3.2068$$
    $$3.0944,2.8368,2.7623,2.6790,2.6558,2.1554,2.0768,1.9787,1.7463,1.2244$$
    $$0.9112,0.6396,0$$
$$\lambda_{12}(8)= 10.3923,\overbrace{3.4641}^{4},\overbrace{3.1623}^{6},\overbrace{2.4495}^{6},\overbrace{0}^{4}$$
$$\lambda_{12}(9)= 10.3923,3.4998,\overbrace{3.4835}^{2},3.3621,\overbrace{3.3535}^{2},\overbrace{3.1446}^{2}$$
$$\overbrace{3}^{2},\overbrace{2.4466}^{2},2.4495,\overbrace{2.2384}^{2},\overbrace{1.7989}^{2},0.6692,\overbrace{0.5012}^{2}$$
$$\lambda_{12}(10)= 10.3923,3.4490,3.4963,\overbrace{2.4549}^{2},\overbrace{3.3425}^{2},3.2578,\overbrace{2.9974}^{2}$$
$$2.6779,2.6114,\overbrace{2.4495}^{2},\overbrace{2.3361}^{2},1.7885,1.5948,1.0896,\overbrace{0.6706}^{2}$$
Another graph is
$$(n,\kappa,\lambda,\mu)=(64,18,2,6)$$.
There are $2$ different eigenvalues of the matrix $A_{12}$.
$$\lambda_{12}(1)=9.4868,\overbrace{3.8730}^{12},\overbrace{0}^{5}$$
$$\lambda_{12}(2)=9.4868,\overbrace{3.8730}^{6},\overbrace{2.6458}^{6},\overbrace{0}^{2}$$
\section{Conclusion}
 The entanglement entropy could be obtained between two  parts in the quantum
networks that their nodes are considered as quantum harmonic oscillators. The Schur
complement method was used to calculate the Schmidt number and entanglement entropy between two
parts of graph. The adjacency matrices of strongly regular graphs were written
 in the stratification basis. we could calculate some important relations for the blocks of adjacency
matrices. Also in four important classes of SRGs, all blocks of adjacency
matrices could be found in terms of the parameters of SRGs analytically. More, the relationship
between size of the boundary of strata and entanglement entropy is obtained in the limit of large coupling.
\\We could develop the quantum algorithms for distinguishing some non-isomorphic pairs of
SRGs, by using the elements of blocks of adjacency matrices in the stratification basis. By this method, we could develop the quantum algorithms for
distinguishing some non-isomorphic pairs of SRGs, by simple way.
\\one expects that the Above methods (stratification basis and the generalized schur complement method) can be used for calculating entanglement entropy in the excited states of quantum harmonic
oscillator and other quantum models.
\\The other aim is that the considered techniques, be generalized to
other kinds of graphs such as association schemes. It is under investigation for some distance regular graphs.
\section*{Appendix}
\appendix
\section{Schur Complement method}
Let $M$ be an $n\times n$ matrix written a as $2\times2$ block
matrix
\begin{equation}
M=\left(%
\begin{array}{cc}
  A & B \\
  C & D \\
\end{array}%
\right)
\end{equation}
where $A$ is a $p\times p$ matrix and $D$ is a $q\times q$ matrix,
with $n = p + q$ (so, $B$ is a $p\times q$  matrix and $C$ is a $q
\times p$ matrix). We can try to solve the linear system
\begin{equation}
M=\left(%
\begin{array}{cc}
  A & B \\
  C & D \\
\end{array}%
\right)\left(%
\begin{array}{c}
  x \\
  y \\
\end{array}%
\right)=\left(%
\begin{array}{c}
  c \\
  d \\
\end{array}%
\right)
\end{equation}
that is $$Ax + By=c$$
\begin{equation}
C x+ Dy = d
\end{equation}
by mimicking Gaussian elimination, that is, assuming that $D$ is
invertible, we first solve for $y$ getting
$$y=D^{-1}(d-Cx)$$
and after substituting this expression for $y$ in the first
equation, we get
$$Ax + B(D^{-1}(d-Cx))=c$$
and
\begin{equation}
(A-BD^{-1}C)x=c-BD^{-1}d
\end{equation}
The invertible matrix, $A-BD^{-1}C$ , is called the Schur
Complement of $D$ in $M$.
\section{Stratification} For an
underlying network $\Gamma$, let $W={\mathcal{C}}^n$ (with
$n=|V|$) be the vector space over $\mathcal{C}$ consisting of
column vectors whose coordinates are indexed by vertex set $V$ of
$\Gamma$, and whose entries are in $\mathcal{C}$. For all
$\beta\in V$, let $\ket{\beta}$ denotes the element of $W$ with a
$1$ in the $\beta$ coordinate and $0$ in all other coordinates. We
observe $\{\ket{\beta} | \beta\in V\}$ is an orthonormal basis for
$W$, but in this basis, $W$ is reducible and can be reduced to
irreducible subspaces $W_i$, $i=0,1,...,d$, i.e.,
\begin{equation}
W=W_0\oplus W_1\oplus...\oplus W_d,
\end{equation}
where, $d$ is diameter of the corresponding association scheme. If
we define
 $\Gamma_i(o)=\{\beta\in V:
(o, \beta)\in R_i\}$ for an arbitrary chosen vertex $o\in V$
(called reference vertex), then, the vertex set $V$ can be written
as disjoint union of $\Gamma_i(\alpha)$, i.e.,
 \begin{equation}\label{asso1}
 V=\bigcup_{i=0}^{d}\Gamma_{i}(\alpha).
 \end{equation}
In fact, the relation (\ref{asso1}) stratifies the network into a
disjoint union of strata (associate classes) $\Gamma_{i}(o)$. With
each stratum $\Gamma_{i}(o)$ one can associate a unit vector
$\ket{\phi_{i}}$ in $W$ (called unit vector of $i$-th stratum)
defined by
\begin{equation}\label{unitv}
\ket{\phi_{i}}=\frac{1}{\sqrt{\kappa_{i}}}\sum_{\alpha\in
\Gamma_{i}(o)}\ket{\alpha},
\end{equation}
where, $\ket{\alpha}$ denotes the eigenket of $\alpha$-th vertex
at the associate class $\Gamma_{i}(o)$ and
$\kappa_i=|\Gamma_{i}(o)|$ is called the $i$-th valency of the
network ($\kappa_i:=p^0_{ii}=|\{\gamma:(o,\gamma)\in
R_i\}|=|\Gamma_{i}(o)|$). For $0\leq i\leq d$, the unit vectors
$\ket{\phi_{i}}$ of Eq.(\ref{unitv}) form a basis for irreducible
submodule of $W$ with maximal dimension denoted by $W_0$. Since
$\{\ket{\phi_{i}}\}_{i=0}^d$ becomes a complete orthonormal basis
of $W_0$, we often write$[15]$
\begin{equation}
W_0=\sum_{i=0}^d\oplus \textbf{C}\ket{\phi_{i}}.
\end{equation}
Let $A_i$ be the adjacency matrix of the underlying network
$\Gamma$. From the action of $A_i$ on reference state
$\ket{\phi_0}$ ($\ket{\phi_0}=\ket{o}$, with $o\in V$ as reference
vertex), we have
\begin{equation}\label{Foc1}
A_i\ket{\phi_0}=\sum_{\beta\in \Gamma_{i}(o)}\ket{\beta}.
\end{equation}
 Then by using (\ref{unitv}) and (\ref{Foc1}),
 we obtain
\begin{equation}\label{Foc2}
A_i\ket{\phi_0}=\sqrt{\kappa_i}\ket{\phi_i}.
\end{equation}

\end{document}